\documentclass[]{acta}
\inputencoding{utf8}
\usepackage{supertabular,lscape,epsfig}
\usepackage{amssymb}
\usepackage{amsmath}
\usepackage[T1]{fontenc}
\usepackage{upgreek}
\usepackage{array}

\usepackage[backend=biber,style=authoryear,maxbibnames=99,giveninits=true,natbib=true]{biblatex}
\defbibenvironment{references}
    {\begin{list}{}{\leftmargin=0pt \itemsep=1em}}
    {\end{list}}
    {}

\newbibmacro*{refjournal}{%
    \iffieldundef{journaltitle}
        {\iffieldundef{booktitle}
            {\printfield{publisher}}
            {\printfield{booktitle}}}
        {\printfield{journaltitle}}%
}

\renewbibmacro*{begentry}{%
    \refitem
        {\printnames{author}}  
        {\printfield{year}}    
        {\usebibmacro{refjournal}} 
        {\printfield{volume}}  
        {\printfield{pages}}   
}

\DeclareBibliographyDriver{article}{\usebibmacro{begentry}\finentry}
\DeclareBibliographyDriver{book}{\usebibmacro{begentry}\finentry}
\DeclareBibliographyDriver{inbook}{\usebibmacro{begentry}\finentry}
\DeclareBibliographyDriver{incollection}{\usebibmacro{begentry}\finentry}
\DeclareBibliographyDriver{inproceedings}{\usebibmacro{begentry}\finentry}
\DeclareBibliographyDriver{misc}{\usebibmacro{begentry}\finentry}

\SetPages{61}{71}

\SetVol{76}{2026}
\newcommand{\ArticleDOI}{10.32023/0001-5237/76.1.4}

\begin{document}
\begin{Titlepage}
\Title{Strange quark star II: the minimal and maximal gravitational mass and the Keplerian configuration}
\Author{Fatemeh Kayanikhoo,}
{Nicolaus Copernicus Astronomical Center of the Polish Academy of Sciences, Bartycka 18, 00-716 Warsaw, Poland\\ email: fatima@camk.edu.pl\\
Research Centre for Computational Physics and Data Processing, Institute of Physics, Silesian University in Opava, Bezru\v{c}ovo n\'am.~13, CZ-746\,01 Opava, Czech Republic}

\Author{Mateusz Kapusta,}
{Astronomical Observatory, University of Warsaw, Al. Ujazdowskie 4, 00-478 Warsaw, Poland\\ email: mr.kapusta@student.uw.edu.pl}
\Author{Miljenko \v{C}emelji\'{c},}
{Nicolaus Copernicus Superior School, College of Astronomy and Natural Sciences, Gregorkiewicza 3, 87-100, Toru\'{n}, Poland\\
Nicolaus Copernicus Astronomical Center of the Polish Academy of Sciences, Bartycka 18, 00-716 Warsaw, Poland\\
Research Centre for Computational Physics and Data Processing, Institute of Physics, Silesian University in Opava, Bezru\v{c}ovo n\'am.~13, CZ-746\,01 Opava, Czech Republic}

\Author{W\l odek Klu\'{z}niak,}
{Nicolaus Copernicus Astronomical Center of the Polish Academy of Sciences, Bartycka 18, 00-716
Warsaw, Poland}

\Author{Leszek Zdunik,}
{Nicolaus Copernicus Astronomical Center of the Polish Academy of Sciences, Bartycka 18, 00-716
Warsaw, Poland}

\Received{Month Day, Year}

\end{Titlepage}

\Abstract{
We employ the MIT bag model with density-dependent bag constant for the equation of state (EOS) to estimate the gravitational mass and Keplerian frequency of rapidly rotating strange quark stars (SQS). 
In a companion paper we discuss the structural parameters of such rotating stars under the influence of strong magnetic fields.
We use the LORENE library to compute the structural parameters at different rotational frequencies in the range of 1100-1300~Hz for a non-magnetized SQS. While there is no minimum limit for the mass of slowly rotating self-bound stars, by computing the maximum rotational frequency, known as the mass-shedding limit, we show that SQS must have a minimum mass to sustain high
rotational frequencies. The mass-shedding frequency in our EOS model is lower than that estimated from the MIT bag model EOS with a fixed bag constant. 
The Keplerian frequency in our model depends linearly on the gravitational mass at the mass-shedding limit (and similarly on the minimum mass) with the slope of 0.08~${\rm kHz}/M_\odot$. We obtain mass limits aligned with the observational data for both the heaviest and the lightest observed pulsars.}
{Pulsars , Compact objects , Relativistic stars }


\section{Introduction}\label{intro}
At extremely high densities, larger than $10^{15}\,{\rm g/cm^3}$, which might be found in the cores of compact objects like neutron stars, the strong force holding quarks together within hadrons can weaken to the point that the quarks are no longer confined within these particles. Instead, they form a dense quark matter phase in which strange quarks can exist. The stability of strange quark matter (SQM) regarding the comparable energy per baryon with the value for $^{56}$Fe of about $930\,{\rm MeV}$ confirms that it might be a stable type of matter \citep{Bodmer_1971,Terazawa_1979,Witten_1984, farhi1984strange,Terazawa_1979a}. 
 
The phase transition of nuclear matter to SQM in extreme conditions in the core of compact objects may lead to the existence of a strange quark star (SQS) or a hybrid star (a neutron star with the core containing SQM). The characteristic parameters of SQSs for the first time were studied independently by \cite{Alcock_1986} and \cite{Haensel_1986}. They studied the stability of SQSs and the stellar parameters of these objects compared to the neutron stars. A decade later, \cite{Weber_1997} discussed quark deconfinement in the core of a neutron star, modelling a core as composed of quarks and gluons and the outer layers of protons and neutrons. The comparison between rapidly rotating quark stars and neutron stars was investigated by \cite{Gondek-Rosinska_2000}. By using the simple form of the MIT bag model for the equation of state (EOS), they computed the maximum rotational frequency for the quark star above 2600~Hz and the maximal gravitational mass of $\sim 1.8\,M_\odot$. 

\cite{Haensel_2009} found the approximate formula independent of EOS to estimate the Keplerian (mass-shedding limit) frequency of both neutron stars and quark stars. \cite{Zhang_2007} studied the millisecond pulsar, \emph{XTE J1739-285} which has a period of 0.8~ms (1250~Hz) and a magnetic field about $4 \times 10^{7}-10^{9}$~G. They obtained the mass and radius of $1.51 \,M_\odot$ and 10.9~km, respectively and suggested that this star is, tentatively, an SQS. \cite{Du2009}, explored submillisecond pulsars as quark stars and found that newborn quark stars could have an initial spin period of approximately 0.1~ms. 
 
While the theoretical studies indicate that SQM is the most stable form of matter under specific extreme conditions, investigations into the potential existence of quark stars have not yet provided direct observational signatures.
The most recent study by \cite{Doroshenko_2022}, based on modelling of the X-ray spectrum and a robust distance estimate from Gaia observations of the supernova remnant \emph{HESS J1731-347}, estimates the mass and radius of the central compact object to be $M_{\rm g}= 0.77 ^{+0.20}_{-0.17}\, M_{\odot}$ and $R= 10.40 ^{+0.86}_{-0.78}\,\mathrm{km}$, respectively. This object is the lightest detected compact object, which might be a quark star with the exotic equation of state. 
  
In this paper, we aim to investigate the minimal and maximal gravitational masses as well as the Keplerian frequency of SQS. We use the MIT bag model with the density-dependent bag constant for EOS, resulting in a frequency compatible with observed pulsars. From our numerical computations of rapidly rotating SQSs, we find the minimum gravitational mass and mass-shedding (Keplerian configuration) of rapidly rotating SQS. From our model, we compute the minimal and maximal gravitational masses consistent with the recent observational data \citep{Dem:2010:Nature:,Zhao:2015:,Abbott:2020:, Doroshenko_2022, Romani_2022}. 

The paper is organised into the following sections: in Section~\ref{NR}, we present our numerical setup; in Section~\ref{EOS}, the EOS model is described; Section~\ref{RES} includes the most important results, and we conclude the study in Section~\ref{CONCL}.

\section{Numerical method}\label{NR}

We described our numerical method in detail in Paper I, of which, for the reader's convenience, we briefly present the essential features. We solve the Einstein field equations within the 3+1 formalism in a stationary, axisymmetric space-time resulting in four elliptic partial differential equations (PDEs) by using the LORENE library \citep[see equations in][]{Bonazzola_1998, Franzon_2017, Chatterjee_2015}, a publicly available software
based on C++. The library\footnote{http://www.lorene.obspm.fr} has been developed to solve PDEs in numerical relativity and the EOS, which is described in the following section.
By employing spectral methods it provides a more accurate approach to solving PDEs than grid-based methods, thus enabling precise calculations. 

We compute the structural parameters of SQS including gravitational mass and radius. Our computational results exhibit a high level of accuracy, with both standard 2-dimensional (GRV2) \citep{GRV2} and 3-dimensional (GRV3) \citep{GRV3} virial identities around $10^{-5}$.
 
We solve the equations for a range of rotational frequencies. For each frequency, the central enthalpy is varied from $0.01~ c^2$ to $0.51~ c^2$ in increments of $0.001~{\rm c}^2$, where $c$ is the speed of light.  

\section{Equation of state} \label{EOS}
\begin{figure*}
 \centering
  \includegraphics[width=218.439bp]{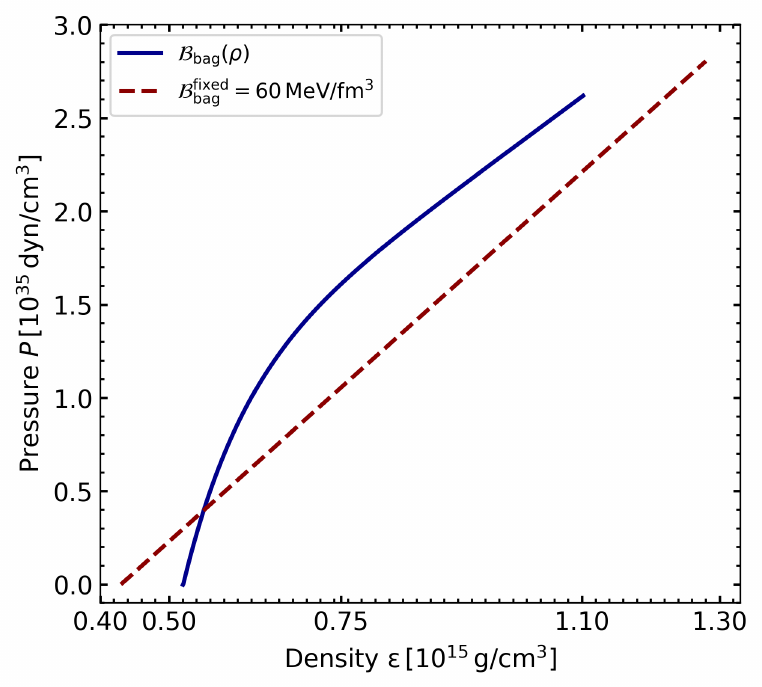}
 \FigCap{The pressure $P$ versus density $\varepsilon$ for two MIT bag models: one with a fixed bag constant $\mathcal{B}^{\rm fixed}_{\rm bag} = 60 \, {\rm MeV/fm^3}$, with red dashed line and one with a density-dependent bag constant $\mathcal{B}_{\rm bag}(\rho)$, with blue solid line.}
 \label{eos}
\end{figure*}
\begin{center}

\begin{table*}[ht!]
\center
\caption{Properties of static SQSs within different EOS models. The slope of EOS is denoted with $a$, the minimum central density is $\varepsilon_0$, and the maximum gravitational mass is $M_{\rm g}^{\rm max}$. For details on the models SS1 and SS2, see \citep[and references therein]{Gondek-Rosinska_2000}}

\resizebox{0.8\textwidth}{!}{

\begin{tabular}{c c c c } 
  \hline
  \textbf{Model} & $\boldsymbol{a}$ & $\boldsymbol{\varepsilon_{0}\, [10^{15}\,{\rm g/cm^3}]}$ & $\boldsymbol{M^{\rm max}_{\rm g}\, [M_\odot]}$  \\
  \hline
  \hline
  $\mathcal{ B}_{\rm bag} (\rho)$ &  0.26 to 1 & 0.52 & 2.35   \\
  $\mathcal{ B}^{\rm fixed}_{\rm bag}$ & 0.33 & 0.43 & 1.96  \\
  SS1 & 0.463 & 1.15 & 1.43  \\
  SS2 & 0.455 & 1.33 & 1.32  \\
  \hline
\end{tabular}}
\label{T1}
\end{table*}
\end{center}

The composition of matter is described by the EOS, which estimates the mass, radius and shape of the object. One of the well-known EOS models for the SQM is the MIT bag model \citep{farhi1984strange}. In this model, quark matter is treated as a collection of non-interacting quarks confined within a finite region of space that is characterized by a parameter known as the bag constant ${\mathcal{B}_{\rm bag}}$. The bag constant is the energy required to create and maintain the bag that confines quarks. In the simple MIT bag model, ${\mathcal{B}_{\rm bag}}$ is considered a fixed value. \cite{Burgio_2002} investigated the bag constant and defined a density-dependent bag model using the experimental results obtained at CERN on the formation of a quark-gluon plasma. The density-dependent bag constant ${\mathcal{B}_{\rm bag}}(\rho)$ is defined as follows, 
\begin{equation}
    {\mathcal{B}_{\rm bag}}(\rho)=\mathcal{B}_{\mathrm{\infty}}+(\mathcal{B}_{0}-\mathcal{B}_{\mathrm{\infty}})e^{-\alpha(\rho/\rho_0)^2},
\end{equation}
where $\rho$ is baryon number density and $\rho_0 = 0.17\, {\rm fm^{-3}}$ is normal nuclear matter density, $\alpha = 0.17 $ and $\mathcal{B}_{0}=\mathcal{B}_{\rm bag}(0)=400~\mathrm{MeV/fm^{3}}$. $\mathcal{B}_{\mathrm{\infty}} = 8.99~\mathrm{MeV/fm^{3}}$ is defined in such a way that the bag constant would be compatible with experimental data (CERN-SPS) \citep{LOCV2000, Burgio_2002}. 

In Fig.~\ref{eos}, the pressure as a function of energy density is shown for the density-dependent MIT bag model, along with the model using a fixed bag constant of $\mathcal{B}^{\rm fixed}_{\rm bag} = 60\, {\rm MeV/fm^3}$ for comparison. In these computations, we consider quarks as free Fermi particles \citep[see details in][]{kayani2020}. The energy density is up to the value corresponding to the maximum mass. 

In $\mathcal{B}^{\rm fixed}_{\rm bag}$ model we linearize the EOS with the function ${ P = a \cdot (\varepsilon - \varepsilon_0)}$ . In this model the slope is $a = 0.33$, and the minimum central density is $\varepsilon_0=0.43\times10^{15}\, {\rm g/cm^3}$.
In $\mathcal{B}_{\rm bag}(\rho)$ model, the slope $a$ is a function of density changing from 1 for the minimum central density less than $0.6\times10^{15}\, {\rm g/cm^3}$ to 0.26 for the larger energy densities. The stiffer EOS in the density-dependent bag constant model results in a larger possible maximum gravitational mass, compared to the model with a fixed bag constant.
In the model with a fixed bag constant, $\mathcal{ B}^{\rm fixed}_{\rm bag}$, for static SQS the maximum gravitational mass is approximately 1.96~${M_{\odot}}$ while in the density-dependent MIT bag model $\mathcal{B}_{\rm bag}(\rho)$ the maximum gravitational mass reaches 2.35~${ M_{\odot}}$.

Two EOS models were investigated by \cite{Gondek-Rosinska_2000}: the SS1 model, with $a = 0.46$ and $\varepsilon_0 = 1.15 \times 10^{15} \, {\rm g/cm^3}$, and the SS2 model, with $a = 0.45$ and $\varepsilon_0 = 1.33 \times 10^{15}\, {\rm g/cm^3}$. In the static SQS, they found the maximum gravitational mass 1.43~$M_\odot$ for the SS1 model and 1.3~$M_\odot$ for the SS2 model. These models could not explain the observed gravitational masses greater than 2~$M_\odot$ while the maximum gravitational mass value computed with model $\mathcal{B}_{\rm bag}(\rho)$ is compatible with the current observational mass $M_{\rm max} \gtrsim 2\,M_\odot$  \citep{Romani_2022, magnetar2015, magnetar2020, Zhao:2015:, Demorest_2010}. 
The parameters for each EOS model in Fig.~\ref{eos}, including those from \cite{Gondek-Rosinska_2000}, are summarized in Table~\ref{T1}.
In the following sections, we use EOS with the density-dependent MIT bag model to study the rotating SQS.
\section{Results and discussion}\label{RES}
In this section, we investigate the approximate formula to estimate the maximum rotational frequency. Then we study the mass and radius of the rapidly rotating SQS. We estimate the minimum mass and the Keplerian configuration in the ${\rm \mathcal{B}(\rho)}$ model. 

\subsection{Maximum rotational frequency}
\begin{figure*}
    \centering
    \includegraphics[width=217.96bp]{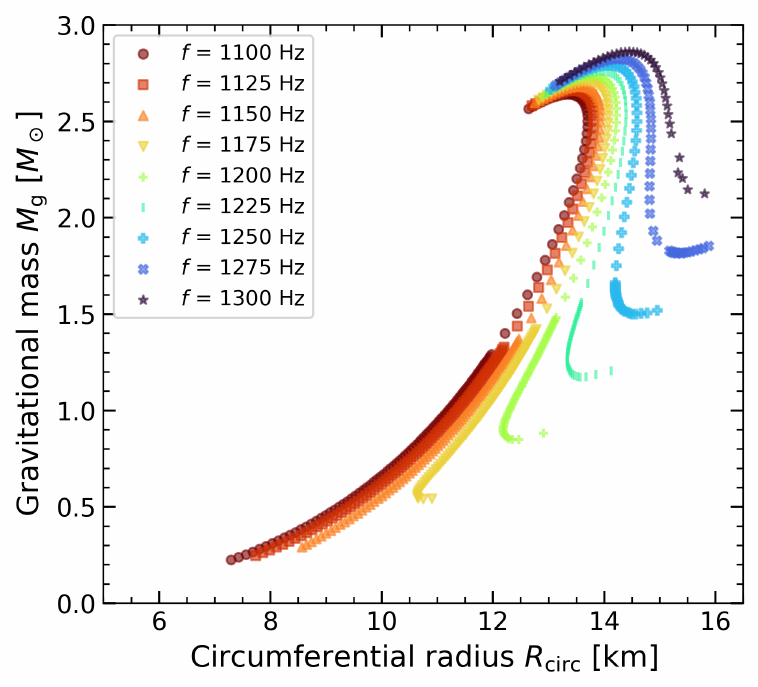}        \caption{The gravitational mass $M_{\rm g}$ versus circumferential radius radius $R_{\rm circ}$ is presented with color-coded filled circles for each frequency. Each marker indicates a computed configuration defined by its central enthalpy, while each color-coded sequence of markers corresponds to a distinct rotational frequency. }
    \label{Mgfrequency}
\end{figure*}

The maximum rotational frequency, $f^{\rm max}$, also known as the mass-shedding limit, denotes the highest spin frequency at which the equatorial surface velocity of the star matches the orbital velocity of a test particle in a circular orbit just above the surface. The maximum rotational frequency of a star configuration can be estimated using the following equation
\begin{equation}
    f^{\rm max} =  C \left(\frac{M_{\rm s}}{M_\odot}\right)^{1/2}\left({\frac{R_{\rm s}}{10\,{\rm km}}}\right)^{-3/2} \label{fmax},
\end{equation}
where ${\rm M_s}$ and ${\rm R_s}$ are the gravitational mass and radius of the non-rotating star in the Newtonian framework, with $C= 1.83\,$kHz. However, the effect of general relativity and the deformation of the star on ${f^{\rm max}}$ are important. It is challenging to determine the relation between the mass and radius of a stationary configuration and the maximum allowed rotational frequency in axisymmetric spacetime due to the star's rotation and curvature effects near massive objects. However, using numerical methods, one can approximate the value of the coefficient $C$ in Eq.~\ref{fmax}. Here we mention two approximations of $C$ and examine it with the data from our model: 
\begin{itemize}
   \item[I)] \cite{Haensel_1995} found this coefficient to be $C=1.22\,$kHz for the maximum mass configuration. 
We use $C = 1.22\,$kHz in the density-dependent bag constant model for the maximum mass configuration and estimate the maximum rotational frequency $\simeq 1.44\,$kHz.

   \item[II)]  \cite{Lattimer_2004} found approximately $C = 1.05\,$kHz for masses not too close to the maximum mass. \cite{Haensel_2009} corrected this value to $C = 1.08\,$kHz and determined constant $C$ for strange quark stars $C_{Q}=1.15\,$kHz. From Eq.\ref{fmax} with $C_Q=1.15$ and the computed configuration in the density-dependent bag constant model with the gravitational mass of 1.4~$M_\odot$ and radius $R=10.65\,$km (see the top-left panel of Figure~2 in the paper I) we obtain the Keplerian frequency 1.24 kHz with excellent agreement with our solution for fast rotating stars. It proves universality of the approximation Eq.\ref{fmax} also for quark stars with the density-dependent Bag constant.

\end{itemize}

\subsection{Minimal and maximal gravitational mass of SQS}
In this section, we use the density-dependent MIT bag model to investigate the gravitational mass limit of SQS.

In Fig.~\ref{Mgfrequency}, is shown the gravitational mass $M_{\rm g}$ versus the circumferential radius ${R_{\rm circ}}$ of SQS at frequencies in the range of 1100~Hz to 1300~Hz, with a spacing of 25~Hz. Each sequence of configurations, specified with a different color, corresponds to a fixed rotational frequency.  
In each sequence, the central enthalpy is monotonic, increasing toward high mass and large radius, up to maximum gravitational mass $M^{\rm max}_{\rm g}$. The maximum of each curve is reached for a constant frequency $f$ and does not set the limit of stable configurations. The boundary of the stable region is defined by the maximum mass at fixed angular momentum and is shifted for a small value to the left with respect to the maximum mass stars presented in Fig.~\ref{Mgfrequency} \citep[see, for example, Fig.3 in][]{Gondek-Rosinska_2000}.

In Fig.~\ref{Mgfrequency}, in the sequences with the frequency $f > 1150$~Hz, there is a minimum mass configuration and a backwards curvature where the equatorial radius increases while the gravitational mass remains nearly constant. The rotation of the star reduces the effective gravitational binding, particularly at the equator, and increases the stellar deformation. As the spin rate rises, the centrifugal force grows, and the equatorial radius expands, weakening the ability of low-mass configurations to maintain hydrostatic equilibrium. The configuration with the smallest gravitational mass in each sequence in Fig.~\ref{Mgfrequency}, $M^{\rm min}_{\rm g}$, is nearly at the mass-shedding limit. The mass-shedding is the last configuration in each sequence, with the frequency higher than 1150~Hz, with a mass $\gtrapprox M^{\rm min}_{\rm g}$ and a larger $R_{\rm circ}$ compared to that corresponding to $M^{\rm min}_{\rm g}$ configuration. For sequences with lower frequencies ($f < 1150$~Hz), within the computational range determined by the selected central enthalpy, neither a pronounced curvature nor the mass-shedding limit is observed; rather, the calculations terminate at the smallest configuration, as depicted in the figure.
We note that, similarly to the case of maximum mass, in the minimum mass configuration, the star does not lose its stability with respect to quasi-radial oscillations---this instability point corresponds to the extremum at fixed $J$.

Fig.~\ref{Mgfrequency} shows that the maximum gravitational mass $M^{\rm max}_{\rm g}$ gradually increases with increasing rotational frequency. $M^{\rm max}_{\rm g}$ changes from 2.55~$M_\odot$ to 2.87~$M_\odot$ as the frequency increases from 1100~Hz to 1300~Hz. The minimum gravitational mass $M^{\rm min}_{\rm g}$ increases from 0.25~$M_\odot$ to 2.15~$M_\odot$ over the same frequency range.  
\begin{figure*}
    \centering
\includegraphics[width=218.141bp]{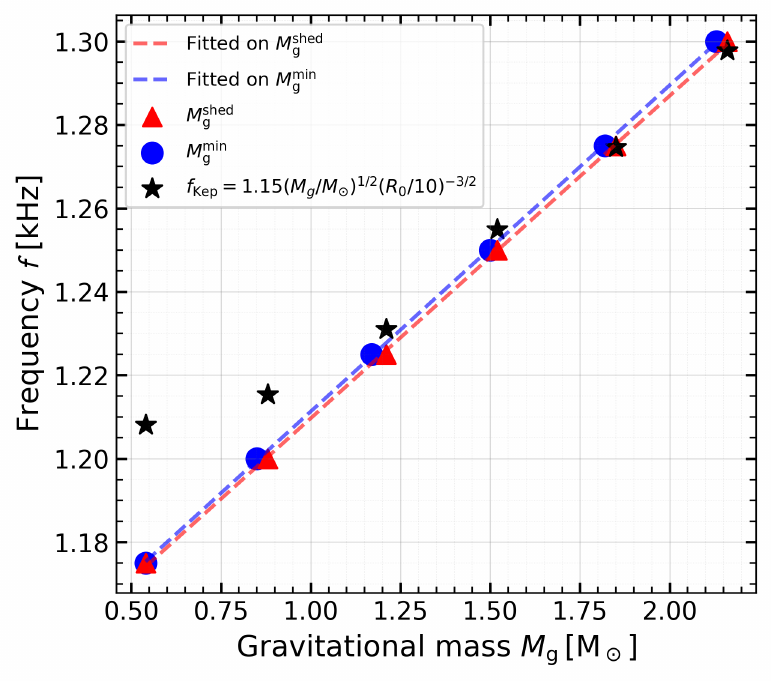}
    \caption{The frequency versus the minimum gravitational mass $M^{\rm min}_{\rm g}$ and mass-shedding $M^{\rm shed}_{\rm g}$ (Keplerian configurations) with blue circles and red triangles, respectively, from our computations are shown along the values from the expression for $f_{\rm Kep}$ by \cite{Haensel_2009} with black stars. The blue and red dashed lines are fitted functions on $M^{\rm min}_{\rm g}$ and $M^{\rm shed}_{\rm g}$, respectively.}
    \label{mass_shedding}
\end{figure*}

The mass-shedding (Keplerian) configuration occurs when the equatorial orbital frequency $f_{\rm eq}$ converges to the rotational frequency of the star. We determine this configuration for each sequence of rotational frequency by calculating the equatorial frequency $f_{\rm eq}$.

Fig.~\ref{mass_shedding}, exhibits the rotational frequency $f$ versus the minimum gravitational mass $M^{\rm min}_{\rm g}$ and mass-shedding $M^{\rm shed}_{\rm g}$. The blue circles indicate the minimum gravitational masses $M^{\rm min}_{\rm g}$, and the red triangles indicate the mass-shedding $M^{\rm shed}_{\rm g}$, both for models with the rotational frequencies $\geq 1175$~Hz. Our results show that both the minimum gravitational mass and the mass-shedding are linear functions of rotational frequency, as indicated by the dotted blue line and dotted red line in the plot, respectively. The fitted function is $f= \Tilde{a} M_{\rm g} + \Tilde{b}$, where $\Tilde{a}$ and $\Tilde{b}$ are constants in the units of $\text{kHz} \,M_\odot^{-1}$ and $\text{kHz}$, respectively. In the case of the minimum gravitational mass, $\Tilde{a} = 0.078$, for the mass-shedding, $\Tilde{a} = 0.077$ and in both functions $\Tilde{b} = 1.13$. The mass-shedding is slightly larger than the minimum mass at each frequency. The difference between the masses of these two configurations increases gradually with increasing rotational frequency. 

A similar linear dependence for self-bound quark stars presented in \cite{Gondek-Rosinska_2000}, shown in Fig.~2, was calculated for a different model, having the maximum mass $M_{\rm max}\simeq 1.35\,M_\odot$ and giving Keplerian frequencies larger than 2000~Hz. However, due to the approximate scaling relations, this frequency is shifted with respect to our model by a factor almost equal to the ratio of maximum mass for these two models and the coefficients $\Tilde{\Tilde{a}}=(df/dM)$ and $\Tilde{\Tilde{b}}=f(0))$ 
scale approximately as $M^{-1}_{\rm max}$ and $M^{-2}_{\rm max}$ respectively, for different models of strange (selfbound) star:
\begin{align}
   \Tilde{\Tilde{a}}= \frac{df}{dM}=0.43\left(\frac{M_{max}}{M_\odot}\right)^{-2}~~~\frac{\rm kHz}{M_\odot}\\
   \Tilde{\Tilde{b}} =f(0) =2.6\left(\frac{M_{max}}{M_\odot}\right)^{-1}~~~{\rm kHz}
\end{align}

To assess the performance of our model, we estimate the Keplerian frequency using the expression given by \citet{Haensel_2009}. They proposed that the Keplerian frequency, $f_{\rm Kep}$, for considered stellar models over a broad range of masses, $0.5 \, M_\odot < M < 0.9 \, M^{\rm stat}_{\rm max}$, can be estimated using the approximate expression,
\begin{equation}
    f_{\rm Kep} = 1.15 \left(\frac{M_{\rm g}}{M_\odot}\right)^{1/2} \left(\frac{R_0}{10\, {\rm km}}\right)^{-3/2}~{\rm kHz}\label{approx}
\end{equation}
where $M_{\rm g}$ is the gravitational mass of the rotating star, $R_0$ is the radius of the non-rotating star of the same mass, and $M^{\rm stat}_{\rm max}$ corresponds to the maximum static gravitational mass of the star. Using this approximate formula, the gravitational mass of the Keplerian configuration for each rotating model, and the corresponding radius in the static model, we estimate the Keplerian frequency.

The approximated $f_{\rm Kep}$ corresponds to the configurations of our models shown by the black stars in Fig.~\ref{mass_shedding}. The data indicate that for configurations with masses $\leq 1.4 \, M_\odot$, the expression gives a slightly higher frequency (less than 2.55\%) compared to our model. The estimated value from the approximate formula (Eq.~\ref{approx}) shown by the black stars converges to the computed frequency in our model (red triangles) for configurations with $M_{\rm g} \geq 1.5 \, M_\odot$.

\section{Summary and conclusions}\label{CONCL}

We developed a model of the rapidly rotating strange quark star using the density-dependent MIT bag model for the equation of state. In this model, the minimum central density is about $0.5 \times 10^{15} \, {\rm g/cm^3}$ and the slope of the EOS is a function of energy density. For densities less than $0.6 \times 10^{15} \, {\rm g/cm^3}$ the slope is equal to one which is larger than that in the MIT bag model with fixed bag constant, and for the densities more than that the slope is about 0.26. This equation of state for the static SQS results in a larger maximum gravitational mass compared to the one computed using the MIT bag model with a fixed bag constant. 
The important results of this work are as follows: 

I) The maximum rotational frequency, $f^{\rm max}$ from Eq.~\ref{fmax}, using the coefficient proposed by \cite{Haensel_1995}, for the static configuration at maximum gravitational mass in our density-dependent MIT bag model is estimated to be 1440~Hz.

II) The coefficient in the estimation formula (Eq.~\ref{fmax}) for the maximum allowed rotational frequency, as determined using our model, is approximately 1150~Hz for configurations with gravitational masses less than the maximum value. This confirms that Eq.~\ref{approx} is valid also for the model with density-dependent bag constant.

III) We have demonstrated that the maximum rotational frequency of strange quark stars shows a linear dependence on both the minimum mass and the mass-shedding configurations, with a slope of approximately $0.08~{\rm kHz}/M_\odot$.

IV) We compared the Keplerian frequencies predicted by our model with those estimated using the approximate formula proposed by \citet{Haensel_2009}. For configurations with gravitational masses $\leq 1.4\,M_\odot$, the formula overestimates the Keplerian frequency by approximately 2.5\%. In configurations with gravitational masses greater than 1.5~$M_\odot$, the Keplerian frequency given by the formula agrees with our model to a fraction of percent.

V) We determine the mass limit of a strange quark star with rotational frequencies between 1100 and 1300 Hz. The maximum gravitational mass is found to vary between 2.55 and 2.78~$M_\odot$, and the minimum gravitational mass for frequencies in the range of 1175 to 1300~Hz spans from 0.54 to 2.13~$M_\odot$. In addition, the maximum gravitational mass for the static configuration is calculated to be 2.35~$M_\odot$. Rotation provides additional centrifugal support against gravitational collapse, allowing more massive configurations to remain stable, so the maximum mass increases with spin.

For low mass self-bound stars, the  Keplerian frequency for a test particle orbiting a spherical star just outside its surface is given by $f_{\rm K} = (2 \pi)^{-1} \sqrt{4 \pi G \varepsilon_0 / 3}$. 
For example, for a density $\varepsilon_0 = 0.5 \times 10^{15}\,{\rm g/cm^3}$, $f_{\rm K} = 1.88\,{\rm kHz}$, which is lower for deformed stars. We note that in the companion paper, we determine that the deformation of a spinning quark star is approximately proportional to the square of its frequency.
To sustain a higher rotational frequency, the star must have a larger average density $\overline{\varepsilon} > \varepsilon_0$, which implies a large mass. 

The presence of a minimum mass at a fixed rotational frequency (or, equivalently, a maximum frequency for given mass that does not correspond to the Keplerian limit) is a characteristic feature of bare strange stars \citep{Gondek-Rosinska_2000}. The inclusion of a crust (with almost negligible mass for strange stars), eliminates this property \citep{Zdunik_2000}.

Our model predicts the gravitational masses of $\geq 2\,M_\odot$
as observed for the pulsars \emph{PSR J1614-2230} ($M=1.908 \pm 0.016M_{\odot}$) and \emph{PSR J0348+0432} ($M= 2.01\pm 0.04 M_{\odot}$)\citep{Dem:2010:Nature:,Zhao:2015:}, and the predicted gravitational mass of $2.6\,M_\odot$, as reported by \cite{Abbott:2020:}, resulting from the compact binary coalescence associated with the recent gravitational-wave event \emph{GW190814} detected by the LIGO/Virgo collaboration.
Our results account for both the heaviest observed pulsar, \emph{PSR J0952-0607} with $M_{\rm g} = 2.35 \pm 0.17 \, M_\odot$ at a rotational frequency of approximately 707~Hz, \citep{Romani_2022},
and the lightest detected compact object, \emph{HESS J1731-347} with $M_{\rm g} = 0.77^{+0.20}_{-0.17}\, M_\odot$ and radius $R = 10.40^{+0.86}_{-0.78}$~km \citep{Doroshenko_2022}. 

A limitation of our model is that the smallest configuration we compute corresponds to a central enthalpy of 0.01~$c^2$, as numerical accuracy for lower values is insufficient. For rotational frequencies below 1175~Hz, the minimum mass and mass-shedding configurations could not be determined due to numerical limitations. 
Furthermore, we note that for rotational frequencies above 1275~Hz, calculating Keplerian configurations becomes increasingly difficult, and at 1300~Hz, our computations encountered numerical failures for some stellar models. In such cases, we use an extrapolation of the equatorial frequency to estimate the mass-shedding configuration. 

As part of our future work, we plan to enhance our numerical methods to accurately determine Keplerian configurations of rapidly rotating, magnetized strange quark stars, as well as to explore alternative equations of state.  
\Acknow{
This project was funded by the Polish NCN grant No. 2019/33/B/ST9/01564. FK is supported by the Polish NCN (Preludium-22 no. 2023/49/N/ST9/01398) and the program Vouchers for Universities in the Moravian-Silesian Region (registration number CZ.10.03.01/00/23\_042/0000390), within the project Accreting Magnetized Neutron Stars. M\v{C} acknowledges the Czech Science Foundation (GA\v{C}R) grant No.~21-06825X and the support by the International Space Science Institute (ISSI) in Bern, which hosted the International Team project \#495 (Feeding the spinning top) with its inspiring discussions. We thank the {\sc LORENE} team for the possibility of using the code.}

\end{document}